\documentclass[twocolumn,pra,aps,showpacs,superscriptaddress,floatfix,showkeys,nofootinbib,10pt]{revtex4-1}
\usepackage{amsfonts,amsmath,amssymb,amsthm}
\usepackage{graphicx}
\usepackage[caption=false]{subfig}
\usepackage{hyperref}
\newcommand{\ket}[1]{ | \, #1 \rangle}
\newcommand{\bra}[1]{ \langle #1 \, |}
\newcommand{\proj}[1]{\ket{#1}\bra{#1}}

\DeclareMathOperator{\Tr}{Tr}

\begin{document}
\title{Device-independent quantum key distribution based on measurement inputs}
\author{Ramij Rahaman}
\email{ramijrahaman@gmail.com}
\affiliation{Department of Mathematics, University of Allahabad, Allahabad 211002, U.P., India}
\affiliation{Institute of Theoretical Physics \& Astrophysics, University of Gda\'{n}sk, 80-952 Gda\'{n}sk, Poland}
\author{Matthew G. Parker}
\email{Matthew.Parker@ii.uib.no}
\affiliation{Department of Informatics, University of Bergen, Post Box-7803, 5020, Bergen, Norway}
\author{Piotr Mironowicz}
\email{piotr.mironowicz@gmail.com}
\affiliation{Department of Algorithms and System Modelling, Faculty of Electronics, Telecommunications and Informatics, Gda\'{n}sk University of Technology, Gda\'{n}sk 80-233, Poland}
\affiliation{National Quantum Information Centre in Gda\'{n}sk, Sopot 81-824, Poland}
\author{Marcin Paw\l{}owski}
\email{dokmpa@univ.gda.pl}
\affiliation{Institute of Theoretical Physics \& Astrophysics, University of Gda\'{n}sk, 80-952 Gda\'{n}sk, Poland}
\date{\today{}}

\begin{abstract}
We provide an analysis of a new family of device independent quantum key distribution (QKD) protocols with several novel features: (a) The bits used for the secret key do not come from the results of the measurements on an entangled state but from the choices of settings; (b) Instead of a single security parameter (a violation of some Bell inequality) a set of them is used to estimate the level of trust in the secrecy of the key. The main advantage of these protocols is a smaller vulnerability to imperfect random number generators made possible by feature (a). We prove the security and the robustness of such protocols. We show that using our method it is possible to construct a QKD protocol which retains its security even if the source of randomness used by communicating parties is strongly biased. As a proof of principle, an explicit example of a protocol based on the Hardy's paradox is presented. Moreover, in the noiseless case, the protocol is secure in a natural way against any type of memory attack, and thus allows to reuse the device in subsequent rounds. We also analyse the robustness of the protocol using semi-definite programming methods. Finally, we present a post-processing method, and observe a paradoxical property that rejecting some random part of the private data can increase the key rate of the protocol.
\end{abstract}

\maketitle

\section{Introduction}
Developments in quantum mechanics lead to emergence of many new research areas including quantum cryptography\cite{QKD}, and quantum computation \cite{QComp}. The goal of quantum information theory is to develop new technologies for information processing that will take us from the traditional classical information age into the age of quantum information. Quantum key distribution \cite{BB84}, the most secure known way for sending secret messages, is a significant achievement in the field of cryptography. It's techniques allow Alice and Bob to establish a shared secret key using an insecure quantum channel and public communication.

Besides the validity of the laws of quantum physics, the security of all QKD schemes relies on some other assumptions. The foremost among them, always present in any such protocol, is that all parties concerned have secure laboratories, \textit{i.e.}, at no stage should there be a leakage of secure classical data from any laboratory. This assumption is crucial and cannot be removed. Another basic assumption is that all players have complete control over their own physical devices, \textit{i.e.}, they have full knowledge over what quantum system their apparatuses use and they also know the exact operation of their measuring devices, \textit{etc.} The goal of the device-independent \cite{MY98} analysis of quantum protocols is to eliminate the latter assumption, \textit{viz.} players can distrust the source of particles and they can also distrust their measuring apparatuses as they might have been prepared by a malicious party.

In 2007, Ac\'in \textit{et al.} \cite{ABGSPS07}, introduced a device independent QKD protocol secure against collective attacks. Earlier questions of a similar type were also addressed by several researchers in different contexts \cite{BHK05,AGM06,SGBMP06}. In 2011, Masanes {\em et al.} \cite{MPA11} provided a more general security scheme based on causally independent measurement processes. The security of all these protocols is undermined as the measurement at any step may depend on the classical or quantum memory of all previous inputs and outputs. Recently secure protocols where device re-use is allowed were introduced \cite{BCK12,VV12}.

In all protocols mentioned above, the parties make measurements on entangled subsystems, check for a violation of some Bell inequality to see if their outcomes are random from the eavesdropper's point of view and, if indeed they are, use them as their secret key. In this manuscript we present a family of protocols which are significantly different. The parties announce their {\it outcomes} and use their choices of measurement {\it settings} for key generation. Our protocol shares this property with the non-device-independent prepare-and-measure SARG04\cite{sarg}.

The potential benefit of flipping the roles of outcomes and settings is that the latter are chosen by the parties using their random number generators (RNGs), which are typically assumed to be perfect, while the former are obtained from measurements on the systems supplied by the eavesdropper. It was shown that even small imperfectness in RNGs are a big threat to QKD \cite{Honza,badRNGqkd}. We demonstrate that they are much a bigger threat to the protocols where the key is obtained from the outcomes than from the settings. More precisely, we take a standard device independent QKD and show that it cannot be secure if the bias of RNGs is greater than 0.1 while our protocol allows for positive key rates far beyond this point. This is the main motivation of our approach.

Obviously, the parties need a way to convince themselves that the correlations they share cannot be classical. Checking for a violation of a Bell inequality is only one possible way of doing so. Another option is to \textit{e.g.} verify the so-called Hardy's paradox \cite{Har92,Fritz12}. There, more than a single security parameter is estimated, which gives the parties more knowledge about the correlations they share. In \cite{LubiePlacki,moreRand,completeStats} this approach has been used to improve the rate of certified randomness.

The main result of this paper is the presentation of a protocol which remains secure, even if the source of randomness used by the parties is strongly biased. Besides that, the protocol serves as a proof of principle that one can use the bits from private random number generators as a key for the device independent QKD. It is shown how a protocol with these properties can be constructed and how its security can be proven. These features may be exploited by some future protocols.

We generalize the results of \cite{MPA11} stating that a condition imposed on a single Bell inequality may certify the randomness of the outcomes. Here we consider the case when there are many parameters used, and the key is formed from the measurement settings with the outcomes made public. What is more, we show that this intrinsically many-valued estimation can be as simple to conduct experimentally, as the standard Bell scenario.

Apart from proving the security of such protocols, we provide a way of using semi-definite programming relaxations to evaluate their key rates. We give explicit numerical results for a protocol basing on the original Hardy's paradox.

\subsection{Organization of the paper}
The organization of our paper is as follows.

We start in Sec. \ref{sec:HardyAndQKD} with recapitulation of the original Hardy's paradox, show the uniqueness of the Hardy state (Sec. \ref{sec:HardyState}), then we describe a QKD protocol and show that in the perfect case it allows for reusing the devices (Sec. \ref{sec:HardyProtocol}).

In Sec.~\ref{sec:biased} we present the main motivation of the paper by discussing the case when the distribution of settings is biased, and compare the presented protocol with other \cite{MPA11} QKD schemes.

We develop methods that allow to analyze the introduced family of protocols when the measurements are causally independent in Sec.~\ref{sec:Methods}. In Sec. \ref{sec:notation} we describe the notation used in the analysis, and in Sec.~\ref{sec:setup} the system configuration. Sec.~\ref{sec:guessing} discusses the definition of the {\em guessing probability of a measurement setting}.

The following Sec.~\ref{sec:sdpRelax} presents a method of evaluation of the guessing probability of a measurement setting using semi-definite programming.

Sec.~\ref{sec:keyrate} discusses the methods of evaluating robustness of the protocol and describes several strategies of post-processing that allow to increase the privacy.

The key rates obtained for these strategies are evaluated, again using semi-definite programming, for the case of Hardy's paradox, in Sec.~\ref{sec:Robustness}.

\section{Hardy's paradox and Quantum Key Distribution}
\label{sec:HardyAndQKD}
In this section we introduce the Hardy's paradox\cite{Har92}, and describe a quantum key distribution protocol based on it.

Consider a physical system consisting of two subsystems shared between two distant parties. The two observers (Alice and Bob) have access to one subsystem each. Both can choose one of two binary measurement settings labeled $0$ and $1$, with outcomes $0$ and $1$. The settings are chosen at random in subsequent runs of the experiment. Settings are denoted by letters $A$ and $B$, while outcomes by $a$ and $b$, for Alice and Bob respectively.

\subsection{The Hardy's state}
\label{sec:HardyState}
The Hardy-type argument starts with the following set of four joint probability conditions for two two-level systems:
\begin{equation}
\label{hardy2q}
\begin{aligned}
P(a=0,b=0|A=0,B=0) \equiv q > 0, \\
P(a=0,b=0|A=1,B=0) = 0, \\
P(a=0,b=0|A=0,B=1) = 0, \\
P(a=1,b=1|A=1,B=1) = 0.
\end{aligned}
\end{equation}
Let us find the set of states $\rho$ for which the conditions of the Hardy-type argument given in \eqref{hardy2q} are satisfied for a given pair of observables. Let us denote the eigenstates of the observable $P=0(1)$ by $\ket{0}_P$ $(\ket{0'}_P)$ and $\ket{1}_P$ $(\ket{1'}_P)$ for the outcome $0$ and $1$ respectively. We now associate a product state with every condition in the test \eqref{hardy2q}, say:
\begin{equation}
\label{prodS}
\begin{aligned}
\ket{\phi_3} &= \ket{0}_A\ket{0}_B, \\
\ket{\phi_2} &= \ket{0'}_A\ket{0}_B, \\
\ket{\phi_1} &= \ket{0}_A\ket{0'}_B, \\
\ket{\phi_0} &= \ket{1'}_A\ket{1'}_B.
\end{aligned}
\end{equation}
Let
\begin{equation}
\label{obser2}
\begin{aligned}
\ket{0'}_P & \equiv \alpha_P \ket{0}_P + \beta_P \ket{1}_P, \text{ and} \\
\ket{1'}_P & \equiv \beta^*_P \ket{0}_P - \alpha^*_P \ket{1}_P,
\end{aligned}
\end{equation}
where $|\alpha_P|^2 + |\beta_P|^2=1$ and $0<|\alpha_P|<1$ for $X=A,B$. The last condition is due to the non-commutativity of $X=0$ and $X=1$.

Let $\mathcal{S}$ be the subspace spanned by the three linearly independent states $\ket{\phi_0}, \ket{\phi_1}$ and $\ket{\phi_2}$ given in \eqref{prodS}. To satisfy the conditions given in \eqref{hardy2q}, $\rho$ has to be confined to a subspace $\mathcal{S}^{\perp}$ of $\mathcal{C}^2\otimes \mathcal{C}^2$, which is orthogonal to $\mathcal{S}$ but not orthogonal to $\ket{\phi_3}$. The dimension of $\mathcal{S}^{\perp}$ is one. Therefore, $\rho$ must be an unique (up to a local unitary) pure two-qubit entangled state, which we denote $\ket{\psi^H}$. Thus, {\it no mixed state of two spin-1/2 particles will satisfy Hardy's argument} \cite{Kar97}. It can also be shown that no two maximally entangled qubit states satisfy Hardy's argument \cite{Har92}.

The four product states $\{\ket{\phi_i}\}_{i=0}^3$ are linearly independent, hence, by the Gram-Schmidt orthogonalization procedure, one can find an orthonormal basis $\{\ket{\phi'_i}\}_{i=0}^3$, in which state $\ket{\phi'_3} = \ket{\psi^H}$ is its last member:
\begin{equation}
\label{hardys2}
\begin{aligned}
\ket{\phi'_0} &= \ket{\phi_0}, \\
\ket{\phi'_i} &= \frac{\ket{\phi_{i}}-\sum^{i-1}_{j=0}\langle \phi'_j|\phi_{i}\rangle\ket{\phi'_j}}{\sqrt{1-\sum^{i-1}_{j=0}|\langle \phi'_j|\phi_{i}\rangle|^2}}, \mbox{~for $i=1,2,3$}.
\end{aligned}
\end{equation}

The probability $q$ in conditions \eqref{hardy2q}, for the Hardy state, $\ket{\psi^H}$, reads
\begin{equation*}
\label{value_q}
q = |\langle \psi^H |\phi_3\rangle|^2 = 1-\sum_{i=0}^2|\langle \phi'_i|\phi_3\rangle|^2=
\frac{|\alpha_A\alpha_B|^2|\beta_A\beta_B|^2}{1-|\alpha_A\alpha_B|^2}.
\end{equation*}
Its maximum is $\frac{5\sqrt{5}-11}{2}$ for $|\alpha_A|=|\alpha_B|=\sqrt{\frac{\sqrt{5}-1}{2}}$ \cite{Jor94}.

\subsection{The protocol}
\label{sec:HardyProtocol}
We consider a scenario in which two distant parties, Alice and Bob, want to generate a secure key. They are allowed to use public classical communication. The QKD protocol proceeds as follows:
\begin{itemize}
\item
S1. In the \textit{initial phase} of the protocol, the two parties obtain pairs of entangled qubits. In each round one of the qubits from each pair is given to Alice, and the other to Bob. Each pair is called a subsystem.
\item
S2. Alice randomly chooses whether to measure $A=0$, or $A=1$ on each of her qubits. Bob does the same by choosing randomly between measurements of $B=0$ and $B=1$. Parties repeat such measurements on all subsystems, and collect statistics. In each run, labeled by $i$, they write down the chosen observables, $A_i$ and $B_i$ respectively, together with the obtained results, $a_i$ and $b_i$.
\item
S3. {\em Check for eavesdropping:} For some randomly selected runs, Alice and Bob both announce their measurement choices ($A_i$ and $B_i$) and the corresponding outcomes ($a_i$ and $b_i$). Alice and Bob publicly compare their announced measurement choices in order to estimate security parameters. For this reason this phase is called the \textit{estimation phase}.
\item
S4. For the remaining runs, Alice and Bob announce only their measurement outcomes, not their bases. Next, to generate their key, they select only those runs for which both of them got outcome $0$. (Alice and Bob ignore those unrevealed pairs that do not have outcomes on both sides equal to $0$, so they are working on some subset of the states.)
\item
S5. For each run with outcomes $0$, they assign a bit value according to their settings.

If the pairs of entangled qubits emitted by the shared source are all perfect copies of the two-qubit `Hardy' states $\ket{\psi^H}$ given by Eqs (\ref{hardys2}) the assigned bit values will be perfectly correlated due to (\ref{hardy2q}). That is, in the ideal case they generate the same key. In the noisy case, \textit{key reconciliation} is required.
\end{itemize}
Device-independent approach allows to quantify all possible interventions of the eavesdropper. These may include influencing the internal working of the devices used by Alice and Bob, \textit{e.g.} by establishing any type of correlation, by coupling to the state, by emitting a different state, or by using measurement settings different from those specified by the protocol.

As mentioned earlier, the ideal Hardy test \eqref{hardy2q} for maximum probability of success $q=\frac{5\sqrt{5}-11}{2}$ is fully device-independent \cite{RZS12} - there is a unique quantum probability distribution associated with this value. The conditions \eqref{hardy2q} assure that both parties got outcome $0$ only if they have chosen the same measurement basis. Then we have
\begin{equation}
\nonumber
\begin{aligned}
0 & < P(a=0,b=0|A=0,B=0) = \frac{5 \sqrt{5} - 11}{2} \\
& < P(a=0,b=0|A=1,B=1) = \sqrt{5} - 2
\end{aligned}
\end{equation}
for a given set of observables and the choice of observable on each side is fully random. The protocol is secure against the most general form of collective memory attacks. Unfortunately this case requires perfect experimental data which is not possible to obtain in practice. The remaining part of this paper analyzes the noisy case.

\section{Biased sources of randomness}
\label{sec:biased}
Before we move to the detailed analysis of the protocol in the case with imperfect experimental data let us present the main motivation of this approach. To this end we will compare the robustness to compromised random number generators in our protocol and the standard one based on CHSH inequality. In both cases we assume that the observed data corresponds to what an experimenter would expect from perfect states and devices.

A common assumption in QKD states that the source of randomness is perfect, meaning that settings are \textit{i.i.d.} with a probability distribution defined by numbers
\begin{equation}
\label{perfectDistribution}
\begin{aligned}
\mathbb{P}&_{perfect}(A,B) = \\
&(P(0,0) = p_{A} \cdot p_{B}, P(0,1) = p_{A} \cdot (1 - p_{B}), \\
&P(1,0) = (1 - p_{A}) \cdot p_{B}, P(1,1) = (1 - p_{A}) \cdot (1 - p_{B}))
\end{aligned}
\end{equation}
with $p_{A} = p_{B} = \frac{1}{2}$ for the uniform probability distribution. In Sec.~\ref{sec:nonuniform} we introduce a \textit{nonuniform} probability distribution with and with $p_{A} = p_{B} = \frac{1}{2} (\sqrt{5} - 1)$.

In this section we consider the case in which the average probability distribution of the source of randomness is given by Eq. \eqref{perfectDistribution}, but in particular runs, the probability distribution is biased in a way known to eavesdropper. For the sake of simplicity we consider biases modeled by changing the parameters $p_{A}$ and $p_{B}$ to $p_{A} \pm \epsilon$ and $p_{B} \pm \epsilon$, respectively, for given $\epsilon$, which gives four possible biased distributions, $\left(\mathbb{P}_{biased,i}(A,B)\right)_{i=1,2,3,4}$.

It is easy to see that the average distribution \eqref{perfectDistribution} can be obtained only if the proportions of all biased distributions in the total number of runs are equal.

Note that if we know only the average distribution given by Eq. \eqref{perfectDistribution}, then for runs with a particular biased distribution $\mathbb{P}_{biased,i}(A,B)$, the observed conditional probabilities are under- or overestimated, \textit{viz.}
\begin{equation}
\label{underOverEstimated}
P_{observed}(a,b|A,B) = P_{actual}(a,b|A,B) \frac{P_{biased,i}(A,B)}{P_{perfect}(A,B)}.
\end{equation}

\begin{figure*}[!htbp]
\centering
\includegraphics[width=0.48\textwidth]{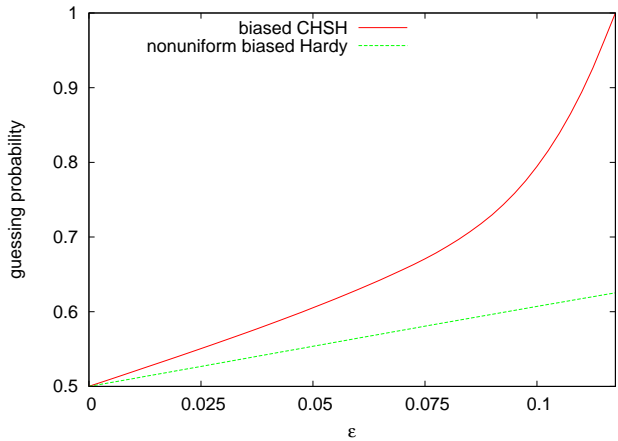}
\caption{(Color on-line) Comparison of guessing probabilities of key values certified with the protocols using Hardy paradox and CHSH inequality in case when the distribution of settings is biased. $\epsilon$ refers to the bias defined in Sec.~\ref{sec:biased}. \label{fig:CHSHvsHardy}}
\end{figure*}

Let us consider the case without noise described by Eq. \eqref{hardy2q} with
\begin{equation}
P(a=0,b=0|A=0,B=0) = q = \frac{5 \sqrt{5} - 11}{2}, \nonumber
\end{equation}
and thus with
\begin{equation}
P(a=0,b=0|A=1,B=1) = \tilde{q} = \sqrt{5} - 2,
\end{equation}
and with nonuniform distribution of settings. Then, for a given biased distribution $\mathbb{P}_{biased,i}(A,B)$, the probability that the generated key is $0$, $P_{i,key=0}$and is given by (\textit{cf.} Eq. \eqref{guessingBayes0})
\begin{equation}
P_{i,key=0} \equiv \frac{q P_{biased,i}(0,0)}{q P_{biased,i}(0,0) + \tilde{q} P_{biased,i}(1,1)}. \nonumber
\end{equation}
The guessing probability is given by $P_{guess,i} = \max(P_{i,key=0}, 1 - P_{i,key=0})$, since the eavesdropper tries to guess the more probable key value. To obtain the average guessing probability this expression has to be averaged over all four possible biased probability distributions, namely
\begin{equation}
\sum_{i=1}^4 \frac{1}{4} P_{guess,i}.
\end{equation}
Guessing probabilities for different $\epsilon$ with nonuniform distribution of settings are shown in Fig.~\ref{fig:CHSHvsHardy}.

In order to compare the efficiency of the presented protocol with other QKD protocols, we use the method of \cite{NPA1,NPA2} to evaluate the guessing probability of the outcomes of Alice certified by the maximal violation the CHSH inequality with uniform, but biased, distribution of settings. We consider the bound on the guessing probability implied by the observed value of $2 \sqrt{2}$ of the following expression
\begin{equation}
\nonumber
\begin{aligned}
4 & \left( P_{biased,i}(0,0) C(0,0) + P_{biased,i}(0,1) C(0,1) \right. + \\
& \left. P_{biased,i}(1,0) C(1,0) - P_{biased,i}(1,1) C(1,1) \right),
\end{aligned}
\end{equation}
where $4$ is the inverse of the uniform probability of each pair of settings in the distribution were unbiased, and $C(A,B)$ is the correlation between outcomes, when the pair of settings $A$ and $B$ is chosen,
\begin{equation}
\nonumber
\begin{aligned}
C(A,B)& \equiv P(0,0|A,B) - P(0,1|A,B) \\
& - P(1,0|A,B) + P(1,1|A,B).
\end{aligned}
\end{equation}
Similarly as in the case of the Hardy protocol, the guessing probability has to be averaged over cases with different biases. The results are shown in fig.~\ref{fig:CHSHvsHardy}. We see that for $\epsilon \approx \frac{1}{10}$ the Hardy protocol is still able to work, whereas the CHSH protocol gives zero key rate.

\section{Methods for analysis of Quantum Key Distribution protocols based on Hardy-like paradoxes}
\label{sec:Methods}
Below we describe the notation used within this paper, the arrangement used in the analysis of the QKD, and give more details about the phases of the QKD protocol.

For the sake of simplicity we consider here a case with perfect RNGs. This can be extended in a natural way to the case with biased probability distributions.
\subsection{Notation and arrangement}
\label{sec:notation}
In the perfect case, we can use the uniqueness of the Hardy state to protect against the collective memory attacks, whereas if the noise occurs we need to assume that the measurements are causally independent, meaning that their operators commute. This is justified by the no-signaling principle if we use many spatially separated measuring devices and perform the measurements on all emitted pairs simultaneously, or if we use a single measuring device that does not have a memory.

We treat successively emitted pairs of particles as separated subsystems. These subsystems together with the subsystem of Eve form one system. We assume that the order of the subsystems is irrelevant. 

Let $L_0$ be a set of labels of pairs of Alice's and Bob's subsystems. For $l \in L_0$ we denote the Hilbert space of the relevant subsystems of Alice and Bob by $\mathcal{H}^\mathcal{A}_l$ and $\mathcal{H}^\mathcal{B}_l$, respectively. The subsystem of Eve lives on a Hilbert space $\mathcal{H}^\mathcal{E}$. We assume that all spaces are finite dimensional. The Hilbert space of the whole system is
\begin{equation}
\label{wholeHilbert}
\mathcal{H} \equiv \mathcal{H}^\mathcal{E} \otimes \prod_{l \in L_0} \mathcal{H}^\mathcal{A}_l \otimes \mathcal{H}^\mathcal{B}_l.
\end{equation}
Vectors on $\mathcal{H}^\mathcal{E}$ are denoted by $\ket{e}^\mathcal{E}$, and on $\mathcal{H}^\mathcal{P}_l$ for $\mathcal{P} \in \{\mathcal{A},\mathcal{B}\}$ by $\ket{e}^\mathcal{P}_l$.

For every pair of subsystems both Alice and Bob perform one of the two measurements, each labeled by either $0$ or $1$. The measurements are binary POVMs denoted by $\tilde{M}^\mathcal{P}_{l,X,x}$, where $\mathcal{P} \in \{\mathcal{A}, \mathcal{B}\}$ denotes the party, $l \in L_0$ denotes the pair of subsystems, $X \in \{0,1\}$ denotes the party's setting, and $x \in \{0, 1\}$ denotes the outcome. The measuring operator $\tilde{M}^\mathcal{P}_{l,X,x}$ acts on $\mathcal{H}^\mathcal{P}_l$.

The natural extension of the operator $\tilde{M}^\mathcal{P}_{l,X,x}$ to the space $\mathcal{H}$ is denoted by $M^\mathcal{P}_{l,X,x}$, and acts with identity operators on spaces different to $\mathcal{H}^\mathcal{P}_l$. From the construction, $M^{\mathcal{P}_1}_{l_1,X_1,x_1}$ commutes with $M^{\mathcal{P}_2}_{l_2,X_2,x_2}$ if $\mathcal{P}_1 \neq \mathcal{P}_2$ or $l_1 \neq l_2$. Recall that $l \in L_0$, and we denote by $a_l$ and $b_l$ the outcomes, and by $A_l$ and $B_l$ the settings, of Alice and Bob, respectively.

Let $S_\mathcal{E}$ be an arbitrary set, and $\{\ket{e}^\mathcal{E}\}_{e \in S_\mathcal{E}}$ be a set of orthogonal states on $\mathcal{H}^\mathcal{E}$. Without any loss of generality, we assume the concerning state in the device independent scenario is of the following form:
\begin{equation}
\label{rhoABE}
\ket{\Phi}^\mathcal{ABE} \equiv \sum_{e \in S_\mathcal{E}} c_e \ket{\phi_e},
\end{equation}
with $c_e \in \mathcal{C}$, $\sum_{e \in S_\mathcal{E}} |c_e|^2 = 1$, where
\begin{equation}
\ket{\phi_e} \equiv \ket{e}^\mathcal{E} \otimes \left( \prod_{l \in L} \ket{e}^\mathcal{A}_l \otimes \ket{e}^\mathcal{B}_l \right), \nonumber
\end{equation}
$\mathcal{P} \in \{\mathcal{A}, \mathcal{B}\}$, $l \in L_0$, and $\ket{e}^\mathcal{P}_l$ is a state on $\mathcal{H}^\mathcal{P}_l$. Eve is allowed to choose the state $\ket{\Phi}^\mathcal{ABE}$, and the measuring operators $\{M^\mathcal{P}_{l,X,x}\}$.

For a subsystem $l$ a conditional probability distribution $\mathbb{P}_l (a,b|A,B) = (P_l(a,b|A,B))_{a,b,A,B}$ can be defined by
\begin{equation}
\label{Pl}
P_l(a,b|A,B) \equiv \Tr \left( \tilde{M}^{\mathcal{A}}_{l,a,A} \tilde{M}^{\mathcal{B}}_{l,b,B} \rho_l \right),
\end{equation}
where $\rho_l$ is the state obtained by tracing all other subsystems in \eqref{rhoABE}.

Let us consider a family of sets of $N_B$ functionals, $\left\{H_k\right\}_{k = 1, \ldots, N_B}$, acting on conditional probability distributions of the form $\mathbb{P}(a,b|A,B) = (P(a,b|A,B))_{a,b,A,B}$ (thus $\mathbb{P}_l$ fits this form). These functionals are defined by a set of values $\{\alpha_{k,a,b,A,B}\}$, with $k = 1, \ldots, N_B$, $a,b,A,B \in \{0,1\}$, and are linear combinations of conditional probabilities of the form
\begin{equation}
\label{Hk}
H_k[\mathbb{P}] = \sum_{a,b,A,B} \alpha_{k,a,b,A,B} P(a,b|A,B).
\end{equation}

From the equation \eqref{hardy2q}, it follows that, in the case of the protocol using original Hardy's paradox, $\alpha_{1,0,0,0,0} = 1$, $\alpha_{2,0,0,1,0} = -1$, $\alpha_{3,0,0,0,1} = -1$, $\alpha_{4,1,1,1,1} = -1$ and the remaining $\alpha_{k,a,b,A,B}$s are equal to $0$.

\subsection{Setups of interest}
\label{sec:setup}
The protocol presented in this paper is device independent, since it relies only on the observed statistics. The main aim is to show that it is possible to prove the security of a key generated out of settings.

In order to illustrate what orders of key rates can be expected to occur in real experiments, we refer to the setup of each subsystem with Hardy's measurements and the following state
\begin{equation}
\label{noise_state}
\rho(\eta) \equiv (1-\eta) \frac{\openone}{4} + \eta \proj{\psi^H}.
\end{equation}

The observed statistics $\mathbb{P}(a,b|A,B)$ do not depend on the distribution of settings, $\mathbb{P}(A,B)$, nevertheless the key rate does. We consider two distributions: uniform and the one described in the section \ref{sec:nonuniform}, referred further as \textit{nonuniform}.

An additional benefit of using nonuniform distribution is the fact that it requires less randomness.

\subsection{The guessing probability of a setting}
\label{sec:guessing}
Let us consider a particular subsystem $l$. We are interested in conditional probabilities of Alice's \textit{settings} $A$, when we know that both Alice and Bob got the outcome $0$, namely $P(A|a=0,b=0)$. These probabilities may be expressed in terms of $\mathbb{P}(a,b|A,B)$ with use of the Bayes rule, as
\begin{subequations}
\begin{equation}
\label{guessingBayes0}
P(A=0|a=0,b=0) = \frac{\sigma}{\sigma+\nu} \text{, and }
\end{equation}
\begin{equation}
\label{guessingBayes1}
P(A=1|a=0,b=0) = \frac{\nu}{\sigma+\nu},
\end{equation}
\end{subequations}
where
\begin{subequations}
\begin{equation}
\label{guessingBayesX}
\begin{aligned}
\sigma \equiv & P(a=0,b=0|A=0,B=0) P(A=0,B=0) + \\
& P(a=0,b=0|A=0,B=1) P(A=0,B=1)
\end{aligned}
\end{equation}
\begin{equation}
\label{guessingBayesY}
\begin{aligned}
\nu \equiv & P(a=0,b=0|A=1,B=0) P(A=1,B=0) + \\
& P(a=0,b=0|A=1,B=1) P(A=1,B=1).
\end{aligned}
\end{equation}
\end{subequations}

Let $\mathbf{h} = \left(h_1, \cdots, h_{N_B} \right)$ denote the values of functionals defined by Eq.~\eqref{Hk} over the probability distribution $\mathbb{P}_l$ defined by Eq.~\eqref{Pl}, so that $h_k = H_k[\mathbb{P}_l]$. For the setup of interest, \eqref{noise_state}, $\mathbf{h}$ is given by the following (\textit{cf.} \eqref{hardy2q}):
\begin{equation}
\label{h_noise}
\begin{aligned}
h_1 = \eta q + \frac{1-\eta}{4}, \\
h_2 = h_3 = h_4 = \frac{1-\eta}{4}.
\end{aligned}
\end{equation}

\begin{figure}[!htbp]
\includegraphics[width=0.45\textwidth]{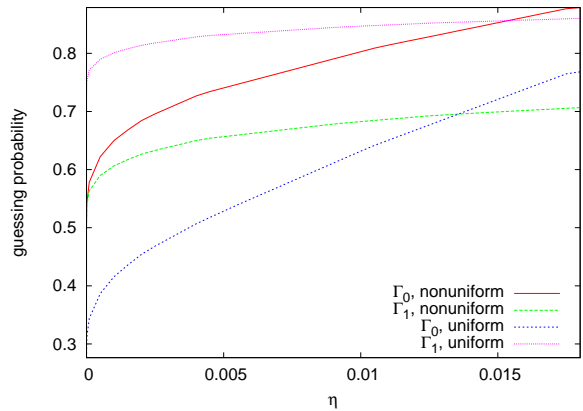}
\caption{(Color on-line) Functions $\Gamma_0(\mathbf{h})$ and $\Gamma_1(\mathbf{h})$ for uniform and nonuniform (see Sec.~\ref{sec:nonuniform}) distribution of settings. These functions give upper bounds for values of $P(A=0|a=0,b=0)$ and $P(A=1|a=0,b=0)$ for $\mathbf{h}$ given by \eqref{h_noise}.\label{fig:gammas}}
\end{figure}

Now, let us ignore the full knowledge about $\mathbb{P}_l$, and consider only the vector $\mathbf{h}$. We introduce two functions, $\Gamma_0(\mathbf{h})$ and $\Gamma_1(\mathbf{h})$, that give upper bounds for values of $P_l (A=0|a=0,b=0)$ and $P_l (A=1|a=0,b=0)$, respectively, allowed by quantum mechanics. Note, that these functions do not depend on $l$, since they do not make any assumptions about the state and the measurements, so they give device-independent bounds. Examples of these functions for $\mathbf{h}$ given by Eq.~\eqref{h_noise} are shown in Fig.~\ref{fig:gammas}.

\section{Semi-definite programming relaxation of the guessing probability}
\label{sec:sdpRelax}
This section describes how to use semi-definite programming \cite{SDP} methods to evaluate upper bounds for functions $\Gamma_0(\mathbf{h})$ and $\Gamma_1(\mathbf{h})$. Expressing them as a semi-definite problem is desired, since such programs may be efficiently treated numerically using the primal-dual interior point algorithm \cite{SeDuMi102,IntPoint}.

In \cite{MPA11} the authors have been able to use a hierarchy of semi-definite programs from \cite{NPA1, NPA2}, called NPA, to find upper bounds for their case. It was possible because they were interested in the probability of guessing the outcome if the setting is known, $P(a|A)$. These probabilities appear directly in the semi-definite programs as problem variables. In our case there is no variable corresponding to $P(A|a,b)$, and therefore we have to find it another way.

Let us consider a subsystem $l$. Without loss of generality, using no signaling principle, we may assume that the eavesdropper performs her measurement with result $e$ before Alice and Bob start the protocol. This does not reduce the generality of the attacks available to the eavesdropper \cite{MPA11}. Moreover, to consider probability distributions allowed for a particular subsystem $l$, we may trace out other subsystems and perform optimization over bipartite states.

To use the NPA method, we introduce functions $\tilde{\Gamma}_0(\mathbf{h})$, and $\tilde{\Gamma}_1(\mathbf{h})$, which give the relevant upper bounds on $P(A|a=0,b=0)$ assuming that the state under consideration is pure. Then $\Gamma_0(\mathbf{h})$, and $\Gamma_1(\mathbf{h})$, are concave hulls of $\tilde{\Gamma}_0(\mathbf{h})$, and $\tilde{\Gamma}_1(\mathbf{h})$, respectively.

We are interested in using the NPA hierarchy in order to calculate $\tilde{\Gamma}_0(\mathbf{h})$. Since the expressions in \eqref{guessingBayes0} and \eqref{guessingBayes1} are not linear in variables occurring in NPA, they cannot by use neither as target, nor as constraint.

To overcome this difficulty, we notice that both $\sigma$ and $\nu$ in \eqref{guessingBayesX} and \eqref{guessingBayesY} are linear in NPA variables. In the general case we need to perform the optimization in two stages. In the first stage we impose the constraints given by the vector $\mathbf{h}$, which can be easily done in the NPA hierarchy and calculate the scope of the allowed values of $\sigma$ for given $\mathbf{h}$. In the second stage we calculate the scope of the allowed values of $\nu$ for given $\mathbf{h}$ and given value of $\sigma$, for some grid of values. This way we obtain a boundary of some region for which it is possible to evaluate the bounds on both \eqref{guessingBayes0} and \eqref{guessingBayes1}.

For Hardy's paradox the optimization is much simpler. In this case $\sigma$ is a function of $\mathbf{h}$. It is easy to see that the expressions \eqref{guessingBayes0} and \eqref{guessingBayes1} achieve their maximal values, if $\nu$ gets it's minimal or maximal value, respectively.

Obviously calculating a function for all possible values of a continuous parameters is impossible. Instead we calculate it only for some grid of values. Now, if we represent the function with set of vectors, each containing the coordinates of a single point together with the value of the function, then the problem of linearly constrained optimization over this function can be solved with linear programming. Examples of such problems are programs stated in Eqs~\eqref{program1} and \eqref{program2} further in this paper.

\section{Evaluating key rates and post-processing strategies}
\label{sec:keyrate}
Here we describe the method of evaluating the key rate achieved by protocols based on Hardy-like paradoxes. We also discuss some post-processing strategies that allow to increase the key rate.

\subsection{Basic case} 
In the simplest case described in Sec.~\ref{sec:HardyProtocol}, the best thing the eavesdropper may do is to maximize his guessing probability, namely $P^{(1)}_{guess}(\mathbf{h})$. Subsystems can be divided into two groups. For states within the first group, the eavesdropper makes a guess that the key value is $0$, and for states from the second group, she guesses the key value $1$.

The probability that a subsystem belongs to the first group is $p_0$. The average values of the Bell observables \eqref{Hk} (or \textit{statistics}) from this group is given by $\mathbf{h}_0$, which allows for guessing $0$ by the eavesdropper with the probability upper bounded by $\Gamma_0(\mathbf{h}_0)$. The remaining part of subsystems (with probability $p_1 = 1 - p_0$) has the statistics given by $\mathbf{h}_1$, and the eavesdropper guesses correctly the key value $1$ with probability not exceeding $\Gamma_1(\mathbf{h}_1)$.

The solution of the following program gives the relevant upper bound for the average guessing probability:
\begin{align}
\label{program1}
\begin{split}
\text{maximize } &\null p_0 \Gamma_0(\mathbf{h}_0) + p_1 \Gamma_1(\mathbf{h}_1) \\
\text{subject to } &\null p_0 \mathbf{h}_0 + p_1 \mathbf{h}_1 = \mathbf{h}, \\
&\null p_0 + p_1 = 1, \\
&\null p_0, p_1 \geq 0.
\end{split}
\end{align}

\begin{figure}[!htbp]
\includegraphics[width=0.45\textwidth]{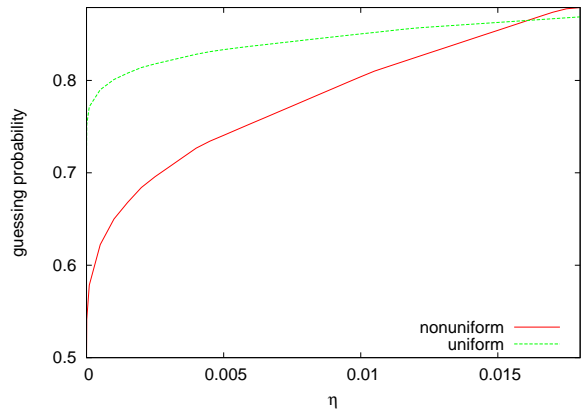}
\caption{(Color on-line) Solutions of the program \eqref{program1} for cases with uniform and nonuniform (see Sec.~\ref{sec:nonuniform}) distribution of settings.\label{fig:guessingProbabilities}}
\end{figure}

Solutions of this program for $\mathbf{h}$ given by \eqref{h_noise} are shown in Fig.~\ref{fig:guessingProbabilities}.

\begin{figure}[!htbp]
\includegraphics[width=0.45\textwidth]{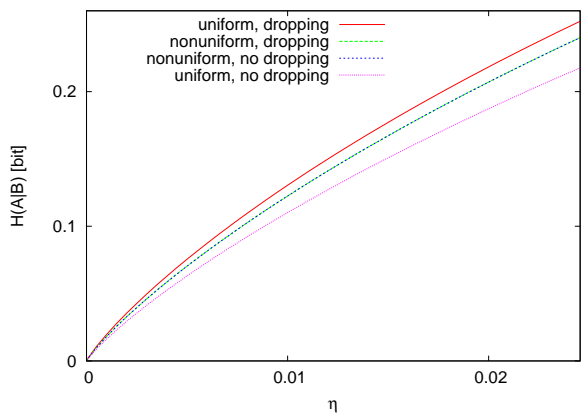}
\caption{(Color on-line) Values of conditional entropies of the setting of Alice given the setting of Bob, $H(A|B)$, if outcomes on both sides were equal to $0$. The cases with uniform and nonuniform distribution of settings, and with and without dropping strategy, are considered. The $\eta$ parameter refers to the state given by Eq. \eqref{noise_state}. In the case with nonuniform distribution, the line referring to use of dropping strategy is slightly above the one without dropping.\label{fig:HABs}}
\end{figure}

The key rate is given by the following formula:
\begin{equation}
K_1 = P(a=0, b=0) \left( -\log_2(P^{(1)}_{guess}(\mathbf{h})) - H(A|B) \right), \nonumber
\end{equation}
where both expressions, $P(a=0, b=0)$ and the conditional entropy $H(A|B)$ can be evaluated from the setup. Examples of conditional entropies for different setups from sec.~\ref{sec:setup} and post-processing strategies, are shown in Fig.~\ref{fig:HABs}.

\subsection{A dropping strategy} 
In a long sequence of $N$ runs, the number of runs with both outcomes equal to $0$ is
\begin{equation}
n \approx P(a=0, b=0) \cdot N. \nonumber
\end{equation}

Eve in $p_0$ fraction of all runs tries to guess that the key value is $0$, and in $p_1 \equiv 1 - p_0$ part of the runs, that it is $1$. The former part of runs gives statistics $\mathbf{h}_0$, and the latter $\mathbf{h}_1$. Since the observed statistics are given by $\mathbf{h}$, we have $\mathbf{h} = p_0 \mathbf{h}_0 + p_1 \mathbf{h}_1$.

Let her success probability in each of these two cases be denoted by $P_0$ and $P_1$, respectively.

Now, let us consider only those runs in which both published outcomes were $0$. Among them the number of runs with the setting of Alice equal to $0$ is
\begin{equation}
\left( p_0 P_0 + p_1 (1 - P_1) \right) n \equiv p^\mathcal{A}_0 n, \nonumber
\end{equation}
and equal to $1$ is
\begin{equation}
\left( p_0 (1 - P_0) + p_1 P_1 \right) n \equiv p^\mathcal{A}_1 n. \nonumber
\end{equation}

If $p^\mathcal{A}_0 < p^\mathcal{A}_1$, then Alice discards (or \textit{drops}) $(p^\mathcal{A}_1 - p^\mathcal{A}_0) \cdot n$ runs with the value $1$. After dropping she has equal number of both values, namely $p^\mathcal{A}_0 \cdot n$ of each. In this situation Eve correctly guesses $p_0 P_0 n$ of runs with the value $0$, and $\frac{p^\mathcal{A}_0}{p^\mathcal{A}_1} p_1 P_1 n$ of runs with the value $1$, so her guessing probability (among those runs that were not dropped) is given by
\begin{equation}
\begin{aligned}
P^{(2)}_{guess} & \equiv \frac{1}{2 p^\mathcal{A}_0 n} \left( p_0 P_0 n + \frac{p^\mathcal{A}_0}{p^\mathcal{A}_1} p_1 P_1 n \right) \\
& = \frac{1}{2} \left( \frac{p_0 P_0}{p^\mathcal{A}_0} + \frac{p_1 P_1}{p^\mathcal{A}_1} \right) \nonumber
\end{aligned}
\end{equation}
The case with $p^\mathcal{A}_0 > p^\mathcal{A}_1$ gives exactly the same formula.

To calculate $P^{(2)}_{guess}(\mathbf{h})$ (as a function of $\mathbf{h}$) \textit{via} a linear optimization, the values $p^\mathcal{A}_0 = P(A=0 | a=0, b=0)$, and $p^\mathcal{A}_1 = P(A=1 | a=0, b=0)$ have to be calculated from the setup. Then the bound on the guessing probability is given by the solution of the following program:
\begin{align}
\label{program2}
\begin{split}
\text{maximize } &\null \frac{1}{2} \left( \frac{1}{p^\mathcal{A}_0} p_0 \Gamma_0(\mathbf{h}_0) + \frac{1}{p^\mathcal{A}_1} p_1 \Gamma_1(\mathbf{h}_1) \right) \\
\text{subject to } &\null p_0 \mathbf{h}_0 + p_1 \mathbf{h}_1 = \mathbf{h}, \\
&\null p_0 + p_1 = 1, \\
&\null p_0, p_1 \geq 0.
\end{split}
\end{align}

The key rate is now given by the following formula:
\begin{equation}
\begin{aligned}
K_2 & = P(a=0, b=0) ( 2 \min(p^\mathcal{A}_0,p^\mathcal{A}_1) ) \\
& \left( -\log_2(P^{(2)}_{guess}(\mathbf{h})) - H(A|B,\text{dropping}) \right)
\end{aligned} \nonumber
\end{equation}

Both expressions, $P(a=0,b=0)$ and the conditional entropy $H(A|B,\text{dropping})$ can be evaluated directly from the setup.

\section{The robustness of Quantum Key Distribution protocols basing on Hardy's paradox}
\label{sec:Robustness}
In this section the method described in Secs~\ref{sec:sdpRelax}~and~\ref{sec:keyrate} is applied to the experimental setup described in the section \ref{sec:setup}. The resulting key rates are shown in Fig.~\ref{fig:keyrates}.

\subsection{Results}
\begin{figure*}[!htbp]
\centering
\includegraphics[width=0.48\textwidth]{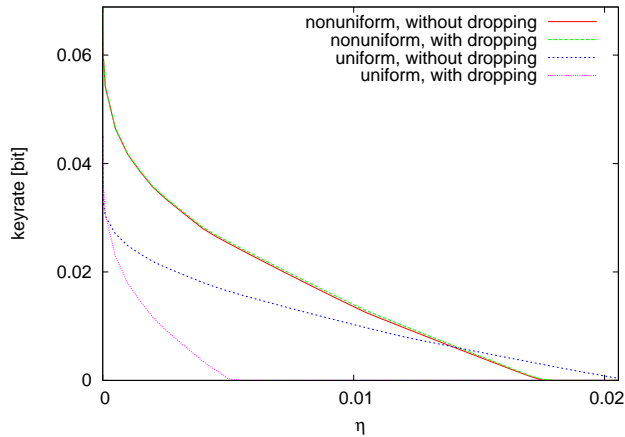}
\caption{(Color on-line) Comparison of key rates in different scenarios.\label{fig:keyrates}}
\end{figure*}

The numerical results concerning the obtained key rates in different situations are shown in Fig.~\ref{fig:keyrates}. The optimal choice of the distribution of settings depends on the value of the noise parameter $\eta$. Although the nonuniform distribution gives better key rates with lower noise, the uniform distribution can be more robust.

In case of nonuniform distribution, the role of the dropping strategy seems to be marginal. This is not surprising, since the aim of that distribution is to make the number of values $0$ and $1$ more or less equal. Similar results refers to conditional entropies (\textit{cf.} Fig.~\ref{fig:HABs}).

A characteristic property of these protocols is the fact, that the use of nonuniform distribution of settings not only requires less randomness, but also in some cases improves the key rate.

\subsection{The case without noise}
The analysis of the case without noise gives an insight to the reason, why the use of the dropping strategy can increase the key rate. It also explains the role of nonuniform distribution of settings.

\subsubsection{The dropping strategy}
In the perfect case with uniform distribution we have $P(a=0,b=0|A=0,B=0) = \frac{5 \sqrt{5} - 11}{2} \approx 0.090167$ and $P(a=0,b=0|A=1,B=1) = \sqrt{5} - 2 \approx 0.236068$, so $P(a=0,b=0|A=0,B=0) < P(a=0,b=0|A=1,B=1)$. Hence, the guessing probability for Eve is higher than $\frac{1}{2}$. The following dropping strategy makes the guessing probability equal to $\frac{1}{2}$.

After performing her measurements, Alice randomly selects only $\frac{P(a=0,b=0|A=0,B=0)}{P(a=0,b=0|A=1,B=1)}$ runs from the total runs with $a=b=0$, where her measurement settings was $A=1$ (in the perfect case, then also $B=1$). Alice sends the list of selected runs to Bob via a public channel.

For this reduced list (from which the key is generated) of runs Alice have equal number of $0$s and $1$s. (In the perfect case they correspond to the same values on the side of Bob.) Hence, the guessing probability is now exactly equal to $\frac{1}{2}$.

We have
\begin{equation}
\begin{aligned} \nonumber
P_\text{not dropped}&(a=0,b=0|A=0,B=0) + \\
& P_\text{not dropped}(a=0,b=0|A=1,B=1) \\
& = 5 \sqrt{5} - 11 \approx 0.180334.
\end{aligned}
\end{equation}
To get the actual ratio of runs that are contained in the key, this should be multiplied by
\begin{equation}
P(A=B=0) = P(A=B=1) = \frac{1}{4}. \nonumber
\end{equation}
Thus the key rate is approximately $\frac{5 \sqrt{5} - 11}{4} \approx 0.04508$.

\subsubsection{Nonuniform distribution of settings}
\label{sec:nonuniform}
Instead of choosing measurement settings with equal probabilities, both Alice and Bob may choose the observables $A=0$ (resp. $B=0$) and $A=1$ (resp. $B=1$) with a ratio $r : 1-r$, for some $r$.

Let us denote for simplicity $P(a=0,b=0|A=0,B=0)=x$ and $P(a=0,b=0|A=1,B=1)=y$. Then to obtain guessing probability equal to $\frac{1}{2}$, the condition for $r$ is
\begin{subequations}
\begin{equation}
x r^2 = y (1-r)^2, \text{ or equivalently}
\end{equation}
\begin{equation}
r = \frac{\sqrt{y}}{\sqrt{x}+\sqrt{y}}.
\end{equation}
\end{subequations}
The key rate is thus $2 x r^2 = 2 x \frac{y}{(\sqrt{x}+\sqrt{y})^2}$.

In the perfect case $x = \frac{5 \sqrt{5} - 11}{2}$ and $y = \sqrt{5} - 2$, so the key rate is $2 x r^2 \approx 0.06888$. The ratio is $r = \frac{1}{2} \left(\sqrt{5} - 1\right) \approx 0.61803$.

\section{Conclusions}
Our paper provides an example of an entirely new class of QKD protocols and provides tools for their analysis.

We have presented a QKD protocol based on Hardy's paradox and analyzed its security in both ideal and noisy scenarios. It has two novel features: (a) The bits used for the secret key do not come from the results of the measurements on an entangled state, but from the choices of settings which are more difficult for an eavesdropper to influence; (b) Instead of a single security parameter a set of them is used to estimate the level of trust in the secrecy of the key, or to construct a certifying observable.

We have shown that these two properties were not chosen by accident. They both make the eavesdropping harder, leading to protocols which can produce a positive amount of shared key even if the biases of the source of randomness are strong. Using more than a single parameter for security provides more information to the parties about the correlations that they share and puts more limits on the eavesdropping strategies.

\section{Acknowledgments}
This work is supported by the Foundation for Polish Science TEAM project (TEAM/2011-8/9/styp7), co-financed by EU European Regional Development Fund and ERC grant QOLAPS (291348), a grant of Ministry of Science and Higher Education of the Republic of Poland IDEAS PLUS (IdP2011 000361), a National Science Centre (NCN) grant 2013/08/M/ST2/00626, and a National Science Centre project Maestro DEC-2011/02/A/ST2/00305. RR also acknowledges partial support by the UGC (University Grants Commission, Govt. of India) Start-Up Grant.

SDP was implemented in the free software package GNU OCTAVE\cite{octave} using the SeDuMi\cite{SeDuMi102,IntPoint} toolbox.

\end{document}